\pdfoutput=1

\documentclass{article}

\usepackage[T1]{fontenc} 
\usepackage[utf8]{inputenc} 
\usepackage{ismir,amsmath,cite,url}
\usepackage{graphicx}
\usepackage{color}
\usepackage{paralist}
\usepackage{units}
\usepackage{microtype}

\usepackage{mwe,tikz}
\usepackage{booktabs}
\usepackage{lineno}

\graphicspath{{../../}}

\title{Representation Learning for the Automatic Indexing of Sound Effects Libraries}

\twoauthors
 {Alison B. Ma} {Music Informatics Group\\Georgia Institute of Technology \\ {\tt \href{mailto:ama67@gatech.edu}{ama67@gatech.edu}}}
 {Alexander Lerch} {Music Informatics Group\\Georgia Institute of Technology \\ {\tt \href{mailto:alexander.lerch@gatech.edu}{alexander.lerch@gatech.edu}}}

\def\authorname{A.B. Ma and A. Lerch}

\usepackage[bookmarks=false,pdfauthor={\authorname},pdfsubject={\papersubject},hidelinks]{hyperref}

\begin{document}

\maketitle
\begin{abstract}
Labeling and maintaining a commercial sound effects library is a time-consuming task exacerbated by databases that continually grow in size and undergo taxonomy updates. Moreover, sound search and taxonomy creation are complicated by non-uniform metadata, an unrelenting problem even with the introduction of a new industry standard, the Universal Category System. To address these problems and overcome dataset-dependent limitations that inhibit the successful training of deep learning models, we pursue representation learning to train generalized embeddings that can be used for a wide variety of sound effects libraries and are a taxonomy-agnostic representation of sound. We show that a task-specific but dataset-independent representation can successfully address data issues such as class imbalance, inconsistent class labels, and insufficient dataset size, outperforming established representations such as OpenL3. Detailed experimental results show the impact of metric learning approaches and different cross-dataset training methods on representational effectiveness.
\end{abstract}

\section{Introduction}\label{sec:introduction}
Sound effects libraries are collections of prerecorded audio assets curated and meant for use by sound designers and editors. In application, a user will search for sounds necessary for their project by querying the library with a computer or other commercial tools for sound search, which helps organize sounds or allow for varying modes of querying the database \cite{yang2020retrieval, lafay2016semantic, elizalde_audiopairbank, vroom}. Examples of typical classes include \textit{ambience}, \textit{foley}, and \textit{sci-fi}. Libraries are often used in game audio and audio post-production applications because access to raw audio assets facilitates efficiency in the creative process, accommodates lacking resources for recording equipment, and alleviates barriers to inconvenient recording locations necessary for particular sounds.

We identify two main problems with sound effects libraries in our work. First, interviews with dataset providers for this research revealed that too much time is spent labeling, re-labeling, and performing quality assurance on databases that continually grow in size and undergo taxonomy updates. One potential solution to this could be the application of machine learning and, more specifically, deep learning for automating sound classification. However, such machine learning models face several particular task-related challenges, such as
\begin{inparaenum}[(i)]
    \item the size of available datasets can dramatically vary in terms of audio data and number of classes, 
    \item data class distributions are often highly imbalanced,
    \item data labels are sometimes of poor quality or inconsistent, and
    \item training dedicated models on every existing sound effects library can be time-consuming.
\end{inparaenum}

Second, non-uniform metadata in sound effects libraries complicate the successful sound search for a user and the creation of a useful taxonomy for a vendor. The recent introduction of the Universal Category System (UCS)\footnote{\href{https://universalcategorysystem.com}{universalcategorysystem.com}, last accessed May 10, 2022.} attempts to address this problem. UCS is an industry-proposed solution to standardize taxonomies designed by and for sound designers and editors. It is designed to be complete and sufficient for sound effects library categorization. However, adopting this standard has been slow; while several sound effects libraries have already converted to this new industry standard, others continue using their own proprietary taxonomies for convenience or preference.

We address problems that arise when using a machine learning approach, training generalized embeddings that can represent any sound effects library and work on any taxonomy for sound classification. To build a powerful learned representation, we investigate two main methods for representation learning with a
\begin{inparaenum}[(i)]
    \item metric-learning and 
    \item cross-dataset training approach.
\end{inparaenum}

Our work has the following main contributions:
\begin{compactitem}
	\item   the introduction of a powerful new representation for use in sound effects libraries trained on relevant, commercially available data,
	\item   the investigation of UCS' generalizability (the industry's first approach to unifying metadata), and
    \item   the presentation of extensive results emphasizing the positive impact of cross-dataset training, an under-researched aspect of representation learning.
\end{compactitem}

The remainder is structured as follows. In Sect.~\ref{subsec:metricLearningLossFunctions}, we analyze if metric learning models will learn a better-structured embedding space and produce higher classification results than a Cross-Entropy (CE) model. In Sect.~\ref{subsec:variableTrainingScenarios}, we see if a cross-dataset training scenario will outperform all other training scenarios and discuss novel UCS-specific findings. In Sect.~\ref{subsec:crossDatasetTrainingScenario}, we evaluate the impact of 3 different cross-dataset training methods. See Sect.~\ref{subsubsec:dataMixing} for data mixing methods, which explores accommodating for encoder bias with data training order, Sect.~\ref{subsubsec:dataReweight}, which investigates accommodating for various dataset characteristics by `Focal' dataset regularization, and Sect.~\ref{subsubsec:ibn}, which experiments with using dataset-independent BatchNorm layers. We compare our best representations against OpenL3 State-of-The-Art (SoTA) deep audio embeddings in Sect.~\ref{subsec:SOTA_comparison}.

\section{Related Work}\label{sec:related_work}
We present an in-depth literature review regarding sound effects libraries and two aspects of representation learning.

\subsection{Sound Effects Libraries}
While there is a plethora of work on sound event classification, there is little work that conducts extensive research for a sound effects library application. Peeters and Reiss conducted sound effects classification by focusing on the discrimination between two classes, ambience and sound effects; audio features were manually selected \cite{peeters_aDeepLearningApproachPostProduction}. Audio tagging work has been done on Freesound \cite{fonseca2018general}, the BBC sound effects library \cite{turnbull2008semantic}, and other online collaborative sound collections \cite{favory_collaborativesoundcollections, font2018sound}. Some have designed features and taxonomies to improve sound classification. Moffat et al.\ tried to address the issue of a non-unified metadata labeling scheme with the Adobe Sound Effects Library.\footnote{\href{https://goo.gl/TzQgsB}{goo.gl/TzQgsB}, last accessed May 10, 2022.} They created a hierarchical taxonomy for sound effects based on the unsupervised learning of sonic attributes, using decision trees to assess audio and label semantic similarity \cite{Moffat2017UnsupervisedTO}. Others that do not target a specific sound effects library application have experimented with psychoacoustic approaches to feature design using actions, material, and mood \cite{lemaitreEvidenceBasicLevel2013b, bonesSoundCategoriesCategory2018b, elizalde_actions, elizalde_audiopairbank, elizalde_never-ending}. Meanwhile, the rest of the work regarding sound effects libraries does not focus on classification but explores sound search, query, and retrieval. Pearce et al.\ investigated user search queries on Freesound \cite{font_freesound_2013}, reducing these descriptors into timbral features to aid sound search \cite{pearceTimbralAttributesSound2017a}. Lafay et al.\ and Zhang et al.\ researched methods to effectively query databases without keywords or 
by vocal imitation \cite{lafay2016semantic, vroom}. Yang et al.\ recently presented a new system for indexing sound effects libraries, evaluating their work on users \cite{yang2020retrieval}.

\subsection{Representation Learning}
Representation learning aims to learn a highly discriminative embedding space that can generalize to various downstream tasks. Two examples of powerful deep audio embeddings are VGGish and OpenL3 \cite{model_cnn_hershey2017cnn, look_listen_learn, look_listen_learn_google}, which have proven to be useful for many music information retrieval (MIR) and cross-modal tasks, e.g., classification, tagging, few-shot learning, and continual learning\cite{few_shot_continual_learning, rethinking_few_shot_learning, gidaris2018dynamic, hierarchical_few_shot}. 

\subsubsection{Metric Learning}
A popular method for representation learning is metric learning, where methods include the Contrastive, Triplet, and Circle loss. Contrastive loss minimizes the distance between similar classes' feature vectors and maximizes the similarity between different classes' feature vectors, training on pairs of positive or negative input samples \cite{fonseca2021unsupervised, spijkervet2021contrastive, khosla}. Triplet loss optimizes both positive and negative inputs simultaneously and adds an additional anchor input in its loss computation \cite{hoffer2015deep, weinberger2009distance, hermans2017defense}. Circle loss tries to improve upon the Triplet loss, featuring more flexible optimization, more definite convergence, and tries to unify the goals of a metric learning and classification loss into one \cite{circle}. Although metric learning losses can be used in a self-supervised \cite{spijkervet2021contrastive, zhang2021transformer, saeed2021contrastive, niizumi2021byol, wang2021towards} or un-supervised \cite{fonseca2021unsupervised} manner, it is common to combine a metric learning loss with a classification loss to better fit a supervised learning problem. Khosla et al.\ have explored training methods to regularize and better structure the embedding space, introducing a Two-Step and Joint-Training method \cite{khosla}. Furthermore, other works have explored the adoption of a contrastive approach to regression \cite{pavan}.

\subsubsection{Cross-Dataset Training}
Furthermore, training on large amounts of diverse data is a crucial aspect of representation learning \cite{model_cnn_hershey2017cnn, look_listen_learn, look_listen_learn_google}. This can be addressed with cross-dataset training; we note many similarities between cross-dataset training and multi-task learning. Though there are scattered works amongst various fields in the deep learning literature that combine multiple different datasets for training, only a few papers have explored proposing methodologies for this. Namely, most work has only been explored in the computer vision domain; there is little work in audio. Working on symbolic music generation, Dong et al. found that stratified sampling mixing alleviated the source imbalance problems that come with combining datasets of various sizes; this worked better than simply concatenating datasets for training \cite{dong2020muspy}. Ranftl et al.\ explored a multi-task learning setup with Pareto-optimal mixing and a multi-objective loss, evaluating success in a zero-shot scenario for a monocular depth estimation task. Their multi-task learning setup and mixing strategy, which ensures that decreasing the loss on one dataset necessitates increasing the loss on another, produced results superior to a naive mixing strategy where they trained on minibatches of equally sampled datasets. They found that this method was better at leveraging the act of adding more datasets for training \cite{crossDataset_mixing}. Wan et al.\ introduced a MultiReader method for a speaker verification task to support training with datasets of different keywords and languages. They regularized the datasets to address insufficient and imbalanced dataset sizes but only experimented with two heterogeneous data sources. Moreover, they do not explicitly illustrate a method of reweighing datasets, resorting to hyperparameter tuning \cite{wan2018generalized}. Lastly, Wang et al. introduced a Dataset-Aware Block, which uses dataset-invariant convolutional layers and dataset-specific BatchNorm layers. They concluded that preserving heterogenous dataset characteristics improves performance \cite{ibn}. To the best of our knowledge, cross-dataset training has not been explored extensively in audio-related tasks.

\section{Method}\label{sec:methodology}
Our best pre-trained model is freely available online. \footnote{\href{https://github.com/alisonbma/aiSFX}{github.com/alisonbma/aiSFX}, last accessed August 11, 2022.}

\subsection{Model Architecture}\label{subsec:modelArchitecture}
\begin{figure}
 \centerline{
 \includegraphics[trim=20 10 10 20, width=0.9\columnwidth]{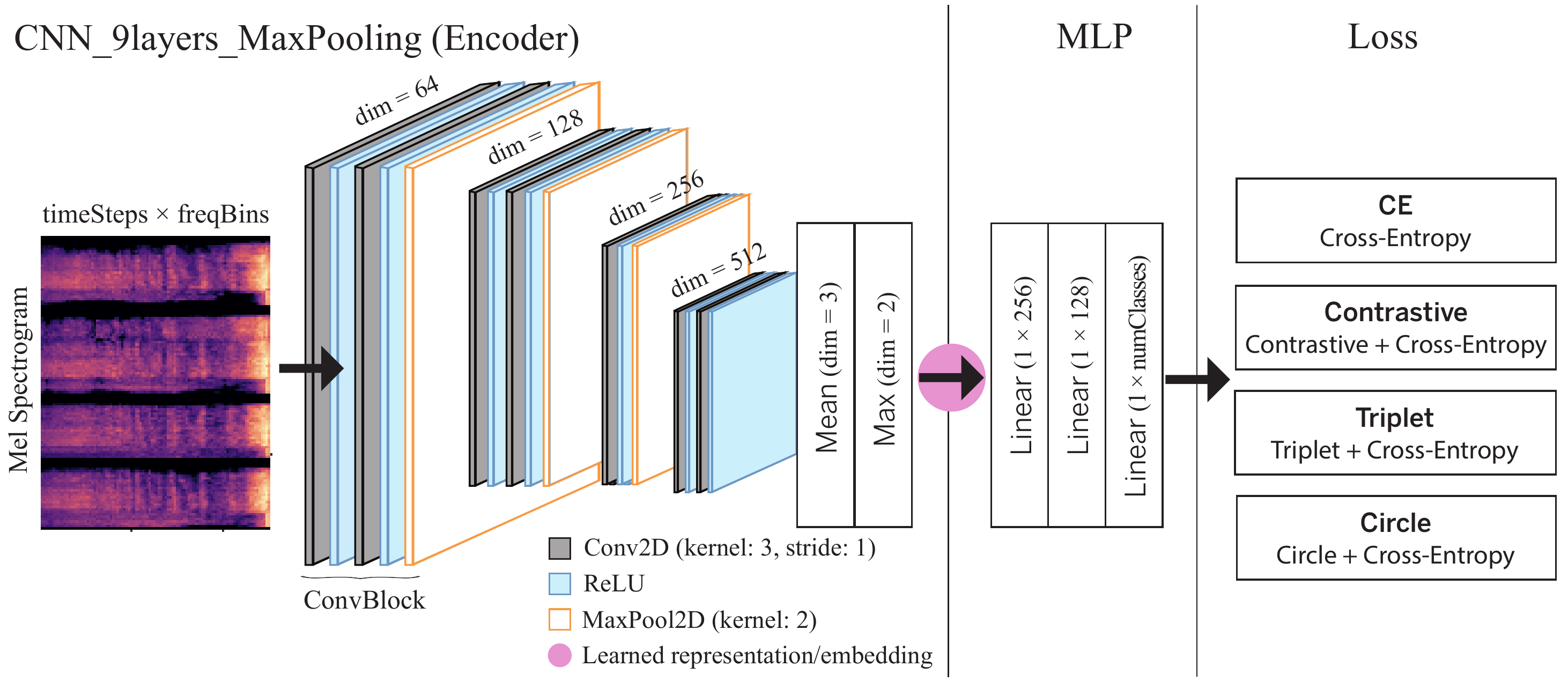}}
 \caption{Network architecture: CNN9-max encoder, MLP classifier head, and loss functions.}
 \label{fig:modelArchitecture}
\end{figure}

Our encoder architecture adapts the Convolutional Neural Network CNN9-max architecture from a DCASE 2019 baseline system for multi-class classification and is displayed in Fig.~\ref{fig:modelArchitecture} \cite{kong2019crosstask}. We pre-train the encoder with one Multi-Layer Perceptron (MLP) classifier head per dataset. After training, we freeze the encoder and pass the representations to a simple Nearest Neighbor classifier to train and test the learned embedding space. Models were trained with Python, PyTorch, and the PyTorch metric learning library.\footnote{\href{https://kevinmusgrave.github.io/pytorch-metric-learning}{kevinmusgrave.github.io/pytorch-metric-learning}, last accessed May 10, 2022.}

\subsection{Input Representation}\label{subsec:inputRepresentation}
All audio files were re-sampled to \unit[44.1]{kHz}, down-mixed, and normalized. Mel-spectrograms were computed with a block size of \unit[46]{ms}, hop size of \unit[23]{ms}, 96 Mel-Bins, and span the audible frequency range \cite{essentia}. We used min-max normalization for spectrograms and z-score standardization for extracted embeddings. Spectrogram input dimensions to the network are (100, 96) and span approximately \unit[2]{s} of audio. This length was determined in pilot experiments.

\subsection{Training Procedure}\label{subsection:modelTraining}
We trained all models with a batch size of 64, the Adam optimizer, and early stopping. Model checkpoints used for inference correspond to the best validation loss. For cross-dataset training, we save the best average validation loss from all datasets. Hyper-parameters for pre-training were found via 20 trials of random search. During pre-training, \unit[2]{s} frames of audio files are randomly sampled whenever a batch is retrieved. We re-shuffle the dataset per epoch. We compute embeddings with 50\% overlap at inference time.

\subsection{Evaluation Metrics}\label{subsec:evaluationMetrics}
The macro F-1 score over classes in a dataset is our predominant metric for evaluating classification performance, as all datasets have imbalanced class distributions. We also monitor the Davies-Bouldin Index (DBI) to evaluate the quality of the embedding space structure, i.e., clustering. The DBI measures the average similarity between a cluster and its most similar cluster.

\section{Datasets \& Taxonomies}\label{sec:datasetsTaxonomies}
\begin{figure}
 \centerline{{
 \includegraphics[trim=15 7 13 7, width=0.9\columnwidth]{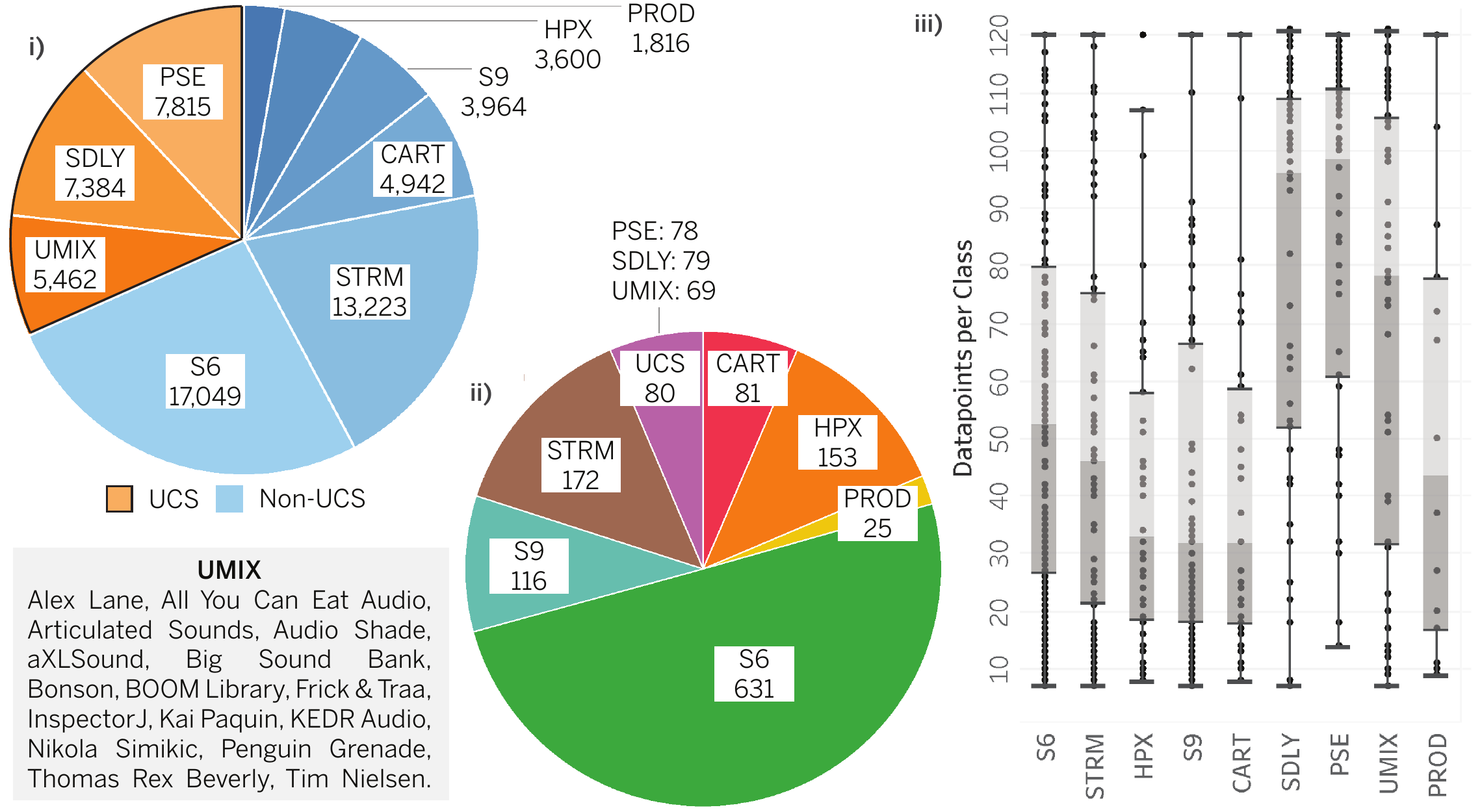}}}
 \caption{Imbalanced UCS \& Non-UCS dataset statistics, (i) Number of audio-files per training dataset, (ii) Number of classes per taxonomy, (iii) Class distribution sorted from most to least classes among training datasets.}
 \label{fig:datasetsTaxonomies}
\end{figure}

We address our goal of training generalized embeddings by training and testing with 9 datasets and 7 taxonomies. We emphasize that we use a diverse assortment of datasets for this research and present corresponding data statistics in Fig.~\ref{fig:datasetsTaxonomies}. We use 3 UCS-compliant datasets: Pro Sound Effects (\textit{PSE}), Soundly (\textit{SDLY}), UCS Mixed (\textit{UMIX}), and 6 Non-UCS datasets from the Sound Ideas library: Cartoon Express (\textit{CART}), HPX Digital (\textit{HPX}), Production Elements (\textit{PROD}), Series 6000 (\textit{S6}), Series 9000 (\textit{S9}), Soundstorm (\textit{STRM}). For UCS-compliant datasets, we use UCSv8.1 \textit{Category} labels as ground truth.

Preliminary experiments with UCS lead us to use experimental subsets of 150 or fewer datapoints per class. We pre-train and evaluate embeddings on the full imbalanced subsets (65k audio files) except in \textit{Cross-Dataset}  training scenarios, where we use class-balanced versions for Non-UCS data. A stratified train validation test split of 8:1:1 was conducted. For UCS-compliant datasets, the fine \textit{CatID} labels determined stratification.

We select \textit{PSE} as the base dataset for \textit{UCS-Transfer} experiments because of its larger number of datapoints and classes, cleaner labels compared to Non-UCS datasets, better generalization when evaluated against other UCS datasets, and more even distribution among both levels of the UCS 2-level class hierarchy.

\section{Experimental Setup}
We list our research questions and hypotheses below using the following experimental variables:
\begin{compactitem}
    \item   \textit{Metric learning loss functions (Fig.~\ref{fig:modelArchitecture}, Sect.~\ref{subsec:exp_rq1})},
    \item   \textit{Variable training scenarios (Sect.~\ref{subsec:exp_rq2})}, and
    \item \textit{Cross-dataset training methods} (Sect.~\ref{subsec:exp_rq3}).
\end{compactitem}

\subsection{RQ1: Impact of Metric Learning}\label{subsec:exp_rq1}
We first ask whether metric learning loss functions that learn distances instead of absolute class labels will improve classification results. We hypothesize that metric learning approaches will learn a more discriminative embedding space compared to CE models and that this better-structured embedding space will improve classification results.

\subsubsection{Training Parameterization}
We train all metric learning models with Khosla's Joint-Training scheme that optimizes a hybrid sum of a metric learning loss function and Cross-Entropy loss as it has been shown to be more effective than a Two-Step training method \cite{niizumi2021byol, NEURIPS2020_d89a66c7, pavan, fan2020multilabel, khosla}. Fig.~\ref{fig:modelArchitecture} illustrates our experiments with the joint metric learning losses
\begin{inparaenum}[(i)]
    \item Contrastive, 
    \item Triplet, and
    \item Circle. 
\end{inparaenum}

The metric learning losses optimize cosine similarity. Positive and negative samples are computed from all possible pairs or triplets between samples in a batch. Accordingly, we find that our batch size is sufficient for training from initial experiments. Model training utilized online mining and the \textit{regularface} embedding regularizer \cite{regularface}.

In addition to re-shuffling the dataset per epoch and randomly selecting \unit[2]{s} frames, we re-sample the dataset so that it randomly selects new datapoints per class per epoch. Here, we equally sample 4 datapoints per class so that a single batch includes 16 different classes; we found minimal difference from weighted sampling.

\subsection{RQ2: Impact of Cross-Dataset Training}\label{subsec:exp_rq2}
Secondly, we ask whether cross-dataset training will improve our best Cross-Entropy and metric learning results in all training scenarios. We hypothesize that a \textit{Cross-Dataset} training scenario will outperform all others, including \textit{UCS-Transfer} (see definition below), improving upon both UCS \& Non-UCS dataset results. Our rationale is that representations generated from \textit{Cross-Dataset} models will have seen a more significant quantity and variety of data from different sound taxonomies, producing more generalized representations. Moreover, we hypothesize that \textit{UCS-Transfer} will yield decent classification results as UCS is a standardized taxonomy for sound effects libraries. We denote \textit{Cross-Dataset} models with the prefix, \textit{X}.

\subsubsection{Training Scenario Definitions}
We conduct experiments with 3 training scenarios:
\begin{inparaenum}[(i)]
    \item   \textit{Within-Dataset} training: Pre-train and evaluate the encoder on the same dataset,
    \item   \textit{UCS-Transfer}: Pre-train the encoder on a UCS-compliant dataset and evaluate on other datasets in a transfer learning scenario, and
    \item   \textit{Cross-Dataset} training: Pre-train the encoder on all datasets and taxonomies, evaluate the encoder on any dataset.
\end{inparaenum}

\subsection{RQ3: Impact of Cross-Dataset Training Methods}\label{subsec:exp_rq3}
Datasets used for training may have varying characteristics. For example, they may use taxonomies of dissimilar scopes and biases and have different dataset sizes, all of which an effective classifier must adapt to. Our final question is whether training scenarios accommodating dataset characteristics will improve classification results. We experiment with 3 methods as introduced in the following subsections,
    \begin{inparaenum}[(i)]
        \item   Data mixing (\textit{X-Sequential}, \textit{X-Joint}),
        \item   `Focal' dataset regularization (\textit{FDR}), and
        \item   Dataset-independent BatchNorm layers (\textit{BN}).
    \end{inparaenum}

\subsubsection{Data Mixing}\label{subsubsec:dataMixing}
We use two variations of data mixing. \textit{Sequential} trains on all datapoints from a single dataset before training on the next, i.e., concatenate datasets, while \textit{Joint} trains on datapoints from all datasets in mixed order, similar to stratified sampling in the literature \cite{dong2020muspy}. We limit mixing for \textit{Joint} so that all datapoints in a batch must correspond to the same dataset. We affirm that backpropagation is called per batch.

\subsubsection{`Focal' Dataset Regularization}\label{subsubsec:dataReweight}
We regularize the datasets by `difficulty' and reweigh them by a dataset's training convergence speed in epochs. Inspired by Focal Loss\cite{wan2018generalized, focal_objectDetection, nada2021multitask} and mining for hard pairs or triplets in metric learning\cite{hermans2017defense, Wu_2017_ICCV, Xuan_2020_WACV, multisimilarity_miner}, easier datasets are down-weighted so that training focuses on difficult datasets.

\begin{equation}\label{dataset_reweighing_factor}
\alpha_d = \frac{1-\beta^{n_e}}{1-\beta}
\end{equation}

We modify the reweighting function shown in Eqn.~\ref{dataset_reweighing_factor} to re-balance the datasets and initialize these weights with $n_e$, the epoch at which a dataset's training set macro F-1 score crosses a threshold of 90\% \cite{reweight_focal}. These weights may be adjusted with hyperparameter $\beta$ to reduce or heighten the difference between dataset weights. $a_d$ is the unnormalized reweighing factor per dataset, $d$. As previously done in the literature, we normalize $\alpha_d$ so that $\sum_{d=1}^{D}\alpha_d=D$, keeping the loss in roughly the same range. $D$ represents the total number of datasets used for pre-training \cite{reweight_focal, focal_objectDetection}.

\subsubsection{Dataset-Independent BatchNorm Layers} \label{subsubsec:ibn}
Similar to selecting the correct classifier head per dataset when pre-training the encoder, we select a dataset's corresponding BatchNorm layers. This is equivalent to turning each ConvBlock in Fig.~\ref{fig:modelArchitecture} into a Dataset-Aware Block \cite{ibn}.

\subsection{RQ4: Comparison to SoTA}
We select OpenL3 features as our reference to SoTA deep audio embeddings because $L^3$-Net is said to ``consistently outperform VGGish and SoundNet on environmental sound classification'' \cite{look_listen_learn}. We use the default OpenL3 parameters as they yield the highest frame-level classification results and select the \textit{Music} content type embeddings for comparison (6144 dimensionality, \unit[0.1]{s} hop size, and 256 Mel Bins) \cite{look_listen_learn}. We aggregate these features to represent \unit[2]{s} of an audio file for consistency with our input representation.

\section{Results}\label{sec:resultsDiscussion}
\begin{figure}
    \centerline{\begin{tikzpicture}[      
            every node/.style={anchor=south west,inner sep=0pt},
            x=1mm, y=1mm,
          ]   
         \node (fig1) at (0,0)
          {{\includegraphics[trim=5 520 925 10, clip, width=0.825\columnwidth]{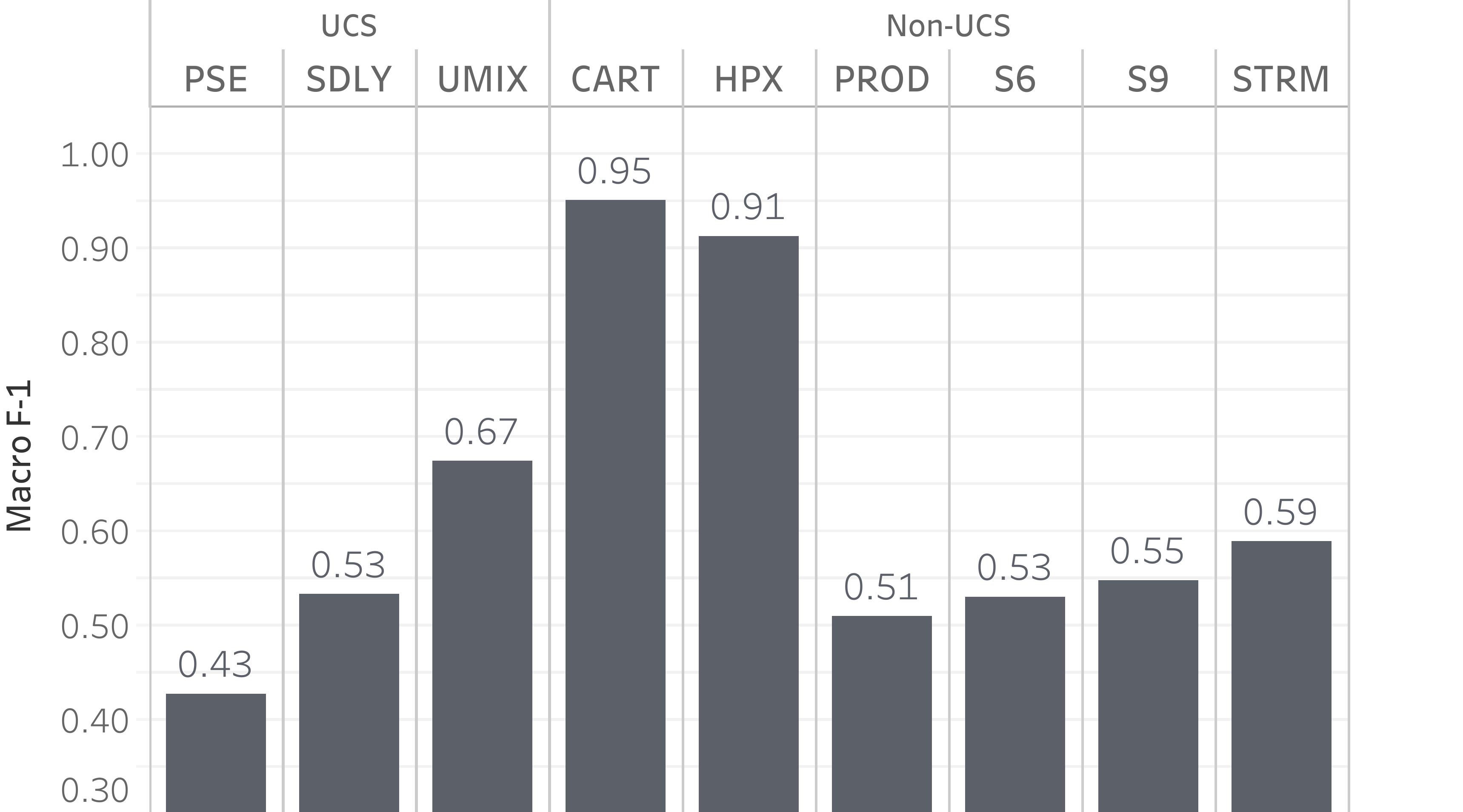}}};
    \end{tikzpicture}}
    \caption{Baseline macro F-1 score classification results using Cross-Entropy loss, CE (\textit{Within-Dataset}).}
    \label{fig:baseline}
\end{figure}

\begin{figure}[!b]
    \centerline{\begin{tikzpicture}[      
            every node/.style={anchor=south west,inner sep=0pt},
            x=1mm, y=1mm,
          ]   
         \node (fig1) at (0,0)
          {{\includegraphics[trim=5 520 925 10, clip, width=0.9\columnwidth]{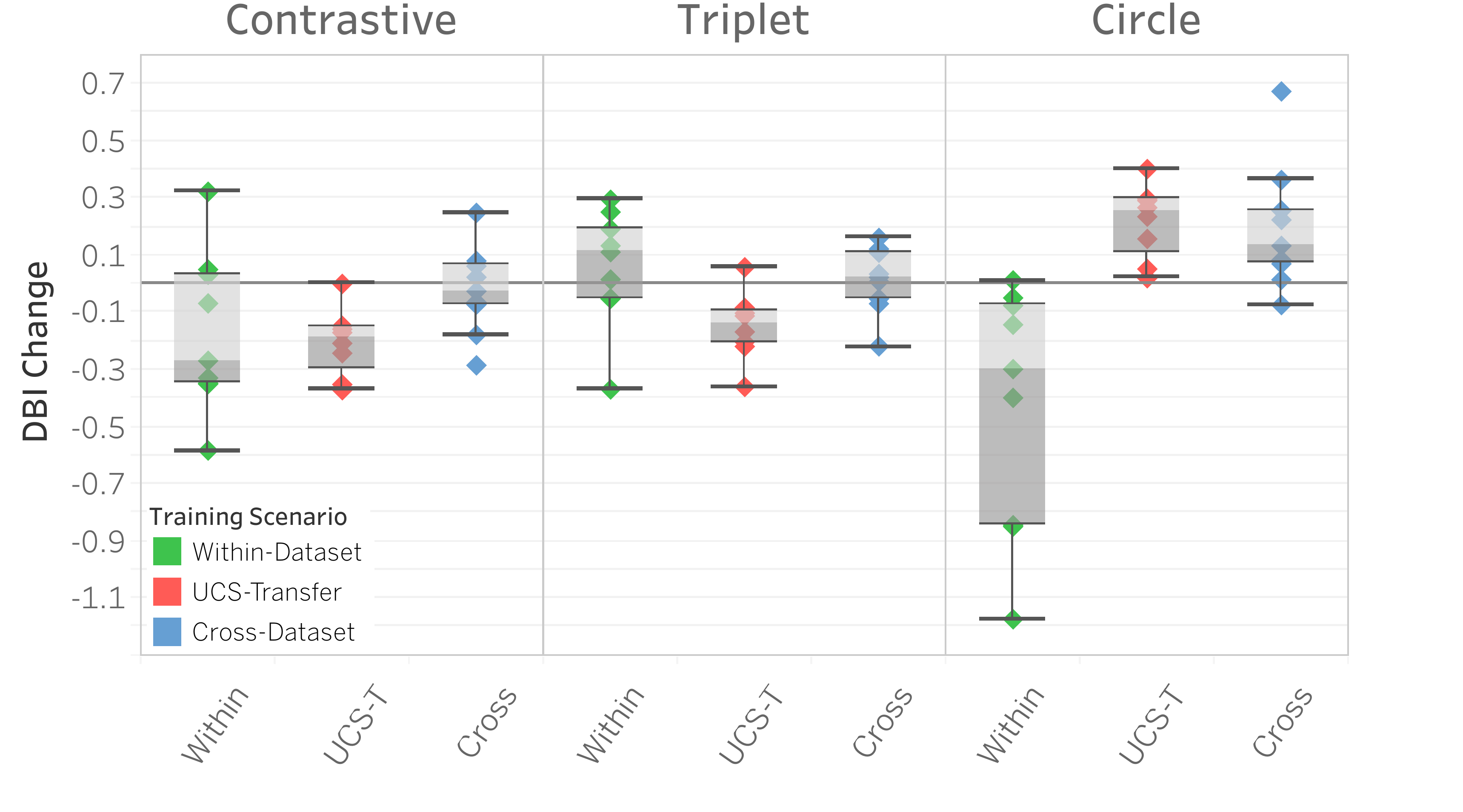}}};
    \end{tikzpicture}}
    \caption{DBI change of metric learning models relative to CE in different training scenarios.}
    \label{fig:metricLossDBI}
\end{figure}

All results are computed at the frame-level, i.e., aggregated over each \unit[2]{s} frame extracted from audio files in the hold-out test set. Boxplot datapoints indicate results for a single dataset. Metric learning plots, Fig.~\ref{fig:metricLossDBI} and \ref{fig:metricLossDBIF1}, only show \textit{X-Sequential} results for the \textit{Cross-Dataset} training scenario.

\subsection{Baseline Results}\label{subsec: baseline_prelim}
Fig.~\ref{fig:baseline} displays baseline classification results where each model is trained and tested on only one dataset, itself. Plots for the following experiments only show the change from this baseline plot.

\subsection{RQ1: Impact of Metric Learning}\label{subsec:metricLearningLossFunctions}

Fig.~\ref{fig:metricLossDBI} shows how metric learning models improve upon the baseline's embedding space structure to different degrees. A lower DBI indicates a better embedding space structure. Contrastive improves the space for most datasets in all training scenarios. Triplet results in a slightly higher average DBI in all but the \textit{UCS-Transfer} training scenario. Circle performs significantly better in a \textit{Within-Dataset} training scenario with an average DBI decrease of $0.43$. However, it performs surprisingly poorly in \textit{UCS-Transfer} \& \textit{Cross-Dataset}, suggesting that it is not ideal for generalization, a prime motivator of representation learning.

\begin{figure}[!b]
    \centerline{\begin{tikzpicture}[      
            every node/.style={anchor=south west,inner sep=0pt},
            x=1mm, y=1mm,
          ]   
         \node (fig1) at (0,0)
          {{\includegraphics[trim=5 520 925 10, clip, width=0.9\columnwidth]{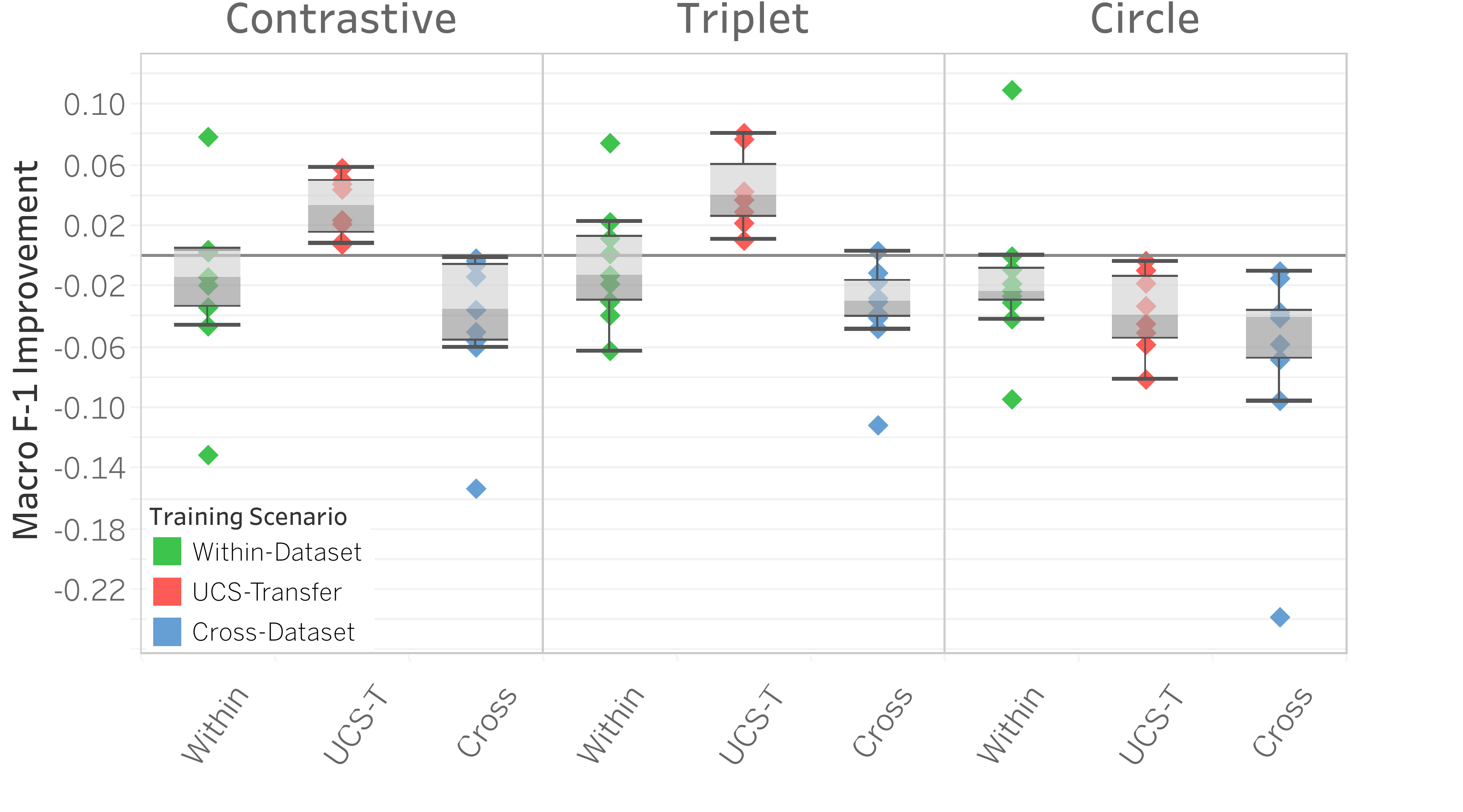}}};
    \end{tikzpicture}}
    \caption{Macro F-1 score improvement of metric learning models relative to CE in different training scenarios.}
    \label{fig:metricLossDBIF1}
\end{figure}

Contrary to our hypothesis, a better-structured embedding space (lower DBI) does not always improve classification results (higher macro F-1), shown in Fig.~\ref{fig:metricLossDBIF1}. We only see that metric learning models improve classification results in the \textit{UCS-Transfer} training scenario. We specifically note that Circle significantly improves the embedding space structure compared to Contrastive in \textit{Within-Dataset}. However, both embedding spaces still yield similar classification performance, an average of $\text{-} 1.75\%$ and $\text{-} 1.50\%$ from CE, respectively. We note other work by Lee et al.\ regarding the similarities between metric learning and classification \cite{metric_vs_classification}.

Focusing on classification results, we identify Triplet as our `best' metric learning loss with a slightly higher average performance than Contrastive. Triplet performs on par with CE in \textit{Within-Dataset} and \textit{Cross-Dataset} (less than $1\%$ difference on average), and has an average of $4.25\%$ improvement from CE in \textit{UCS-Transfer}. Therefore, Contrastive and Circle loss results will be omitted in the following sections.

\subsection{RQ2: Impact of Cross-Dataset Training}\label{subsec:variableTrainingScenarios}
Results for cross-dataset training are shown in Fig.~\ref{fig:trainingScenarioF1Improvement}. Matching our hypothesis, we find that for CE models, \textit{Cross-Dataset} outperforms \textit{Within-Dataset} models for the majority of datasets by approximately $4\text{-}5\%$. We find that Triplet models also tend to improve by $1\text{-}3\%$ except for \textit{X-Joint-BN + Triplet}. This confirms that cross-dataset training indeed leads to better generalization across all sound taxonomies.

\subsubsection{UCS Insights}\label{subsubsec:ucs_insights}
We preface this section with preliminary results on  UCS. We find that the UCS-compliant datasets in this study use the standardized UCS taxonomy in a non-uniform way. Without fine-tuning another classifier head, one cannot use a UCS taxonomy-trained model on other UCS-compliant datasets. This corroborates the need for generalized representations that can adapt to any taxonomy of sound.

\begin{figure}
    \centering{\begin{tikzpicture}[      
            every node/.style={anchor=south west,inner sep=0pt},
            x=1mm, y=1mm,
          ] 
         \node (fig1) at (0,26)
          {\includegraphics[trim=5 475 925 10, clip, width=0.9\columnwidth]{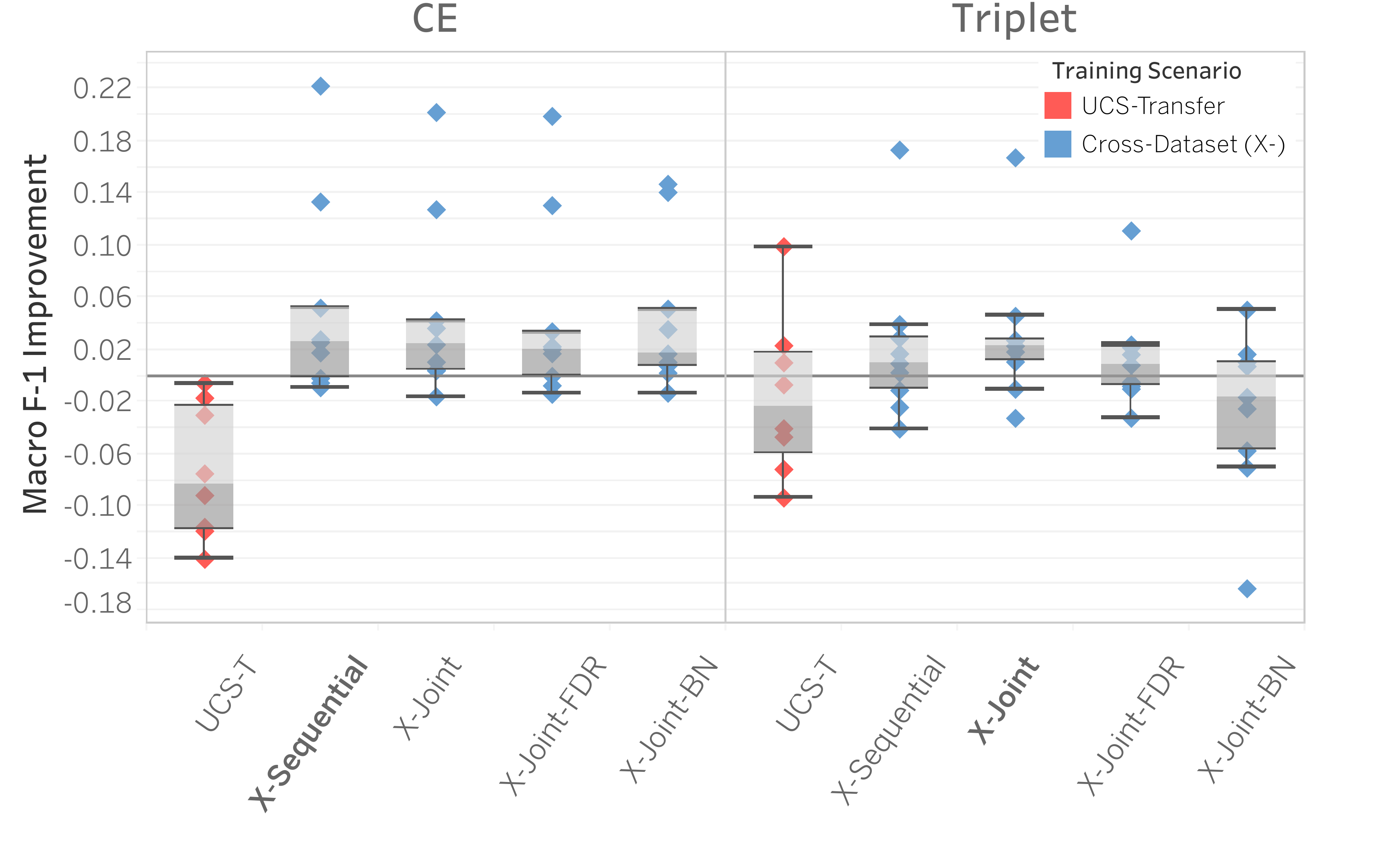}};
    \end{tikzpicture}}
    \caption{Macro F-1 score improvement of various training scenarios from \textit{Within-Dataset}.}
    \label{fig:trainingScenarioF1Improvement}
\end{figure}

With this said, Fig.~\ref{fig:trainingScenarioF1Improvement} also shows that \textit{UCS-Transfer} can achieve results considerably better than random. Although it under-performs \textit{Within-Dataset} training by an average of $7.48\%$ with CE models, we find that there is only an average of $\text{-}1.62\%$ difference from CE with Triplet models. Alongside, we also verify that other UCS-compliant libraries, \textit{SDLY} and \textit{UMIX}, can be used as base sets to achieve decent generalization in a transfer learning scenario.

We note some dataset-specific details for \textit{UCS-Transfer + Triplet} compared to our best \textit{Cross-Dataset} Triplet model, \textit{X-Joint + Triplet}. Both perform similarly (around $1\%$ worse) on Non-UCS \textit{CART, HPX, PROD}, and \textit{STRM} datasets. \textit{UCS-Transfer} performs around $5 \text{-} 6\%$ worse on UCS-compliant \textit{SDLY} and \textit{UMIX}, a substantial improvement from previous results that did not re-train a classifier head. Lastly, it performs around $7\%$ worse on Non-UCS \textit{S6}, likely because \textit{S6} exhibits the highest number of classes amongst all datasets while \textit{PSE} only has 78 classes.

\subsection{RQ3: Impact of Cross-Dataset Training Methods}\label{subsec:crossDatasetTrainingScenario}
We experiment to see if accommodating different dataset characteristics with 3 cross-dataset training methods will further improve results. We compare all cross-dataset training methods to \textit{X-Sequential} in our discussion. 

First, sensitivity to encoder bias and training order does not severely impact results for our task. Data mixing results note an average $1\%$ difference between \textit{X-Sequential} and \textit{X-Joint} in CE and Triplet.

Second, reweighing datasets as proposed in Sect.~\ref{subsubsec:dataReweight} does not lead to observable consistent trends. We show results that use $\beta=0.999$. Looking at difficult datasets, the performance of \textit{SDLY} worsens by 2.96\%, while \textit{STRM} improves by 3.27\% despite having similar convergence speeds and being the second-ranked most `difficult' datasets. We also observe an unfortunate worsening of results for faster-converging datasets, i.e., \textit{HPX} decreases by 1.77\% (Triplet), \textit{S9} decreases by 2.32\% (CE) and $6.20$\% (Triplet).

Third, we verify that independent BatchNorm layers do preserve dataset-specific characteristics. Unlike other work in the literature \cite{ibn}, we find that this may not always be beneficial as it may exacerbate dataset-specific problems. Shown in Fig.~\ref{fig:trainingScenarioF1Improvement}, we specifically note decreases in performance for Non-UCS \textit{HPX, PROD, S6}, and \textit{S9}, which we attribute to messier inconsistent labels and insufficient training data, apparent in Fig.~\ref{fig:datasetsTaxonomies}. This decrease can be quite dramatic with \textit{S9} dropping by $7.53\%$ (CE) and $24.4\%$ (Triplet). Alongside, we find that UCS-compliant \textit{PSE} and \textit{SDLY}, as well as Non-UCS \textit{STRM} experience a $2 \text{-} 3\%$ increase in performance, which we attribute to both the cleaner UCS taxonomy and datasets having sufficient training data.

\subsection{RQ4: Comparison to SoTA}\label{subsec:SOTA_comparison}
\begin{figure}
    \centerline{\begin{tikzpicture}[      
            every node/.style={anchor=south west,inner sep=0pt},
            x=1mm, y=1mm,
          ]   
         \node (fig1) at (0,0)
          {{\includegraphics[trim=5 520 925 10, clip, width=0.9\columnwidth]{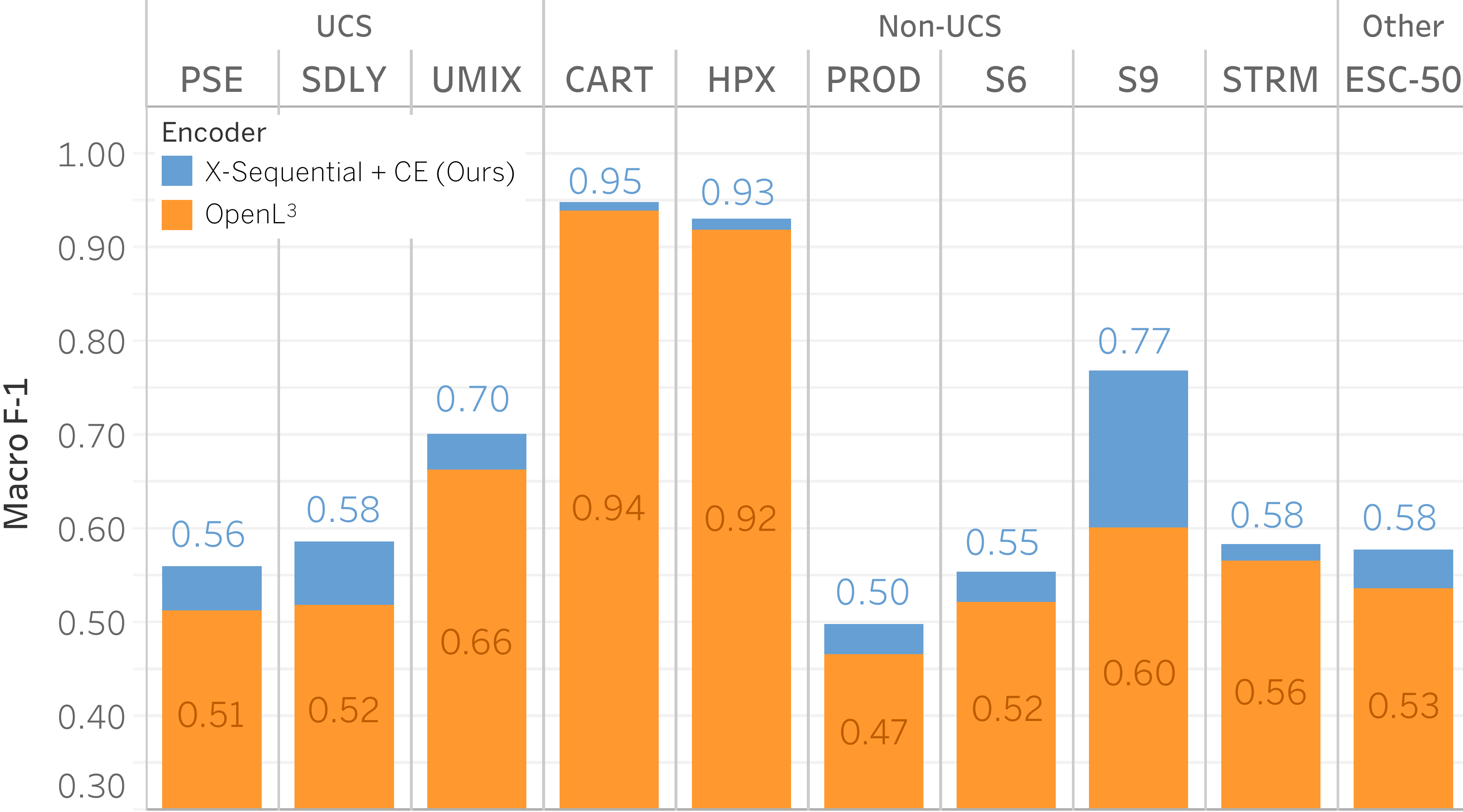}}};
    \end{tikzpicture}}
    \caption{Our best model compared to OpenL3 audio embeddings. Blue bars illustrate the amount in which we outperform OpenL3.}
    \label{fig:compareOpenL3}
\end{figure}

Overall, our work reveals that our best model is \textit{X-Sequential + CE}. We compare \textit{X-Sequential + CE} and OpenL3 in  Fig.~\ref{fig:compareOpenL3} and provide a commonly used benchmark dataset, ESC-50\cite{esc50}, to extend the assessment of our work in a non-task-specific audio application. Unlike the rest of our datasets, we show the 5-fold cross-validation results for ESC-50 as commonly reported in the literature. Ultimately, we observe that our best model outperforms OpenL3 on all datasets. In addition, we note that our model requires fewer resources to achieve its results. While both models have a similar number of parameters (4.7M), our best model is trained on approximately 7x less data (39k audio-files vs.\ 296k videos) \cite{look_listen_learn}.

\section{Conclusion}\label{sec:conclusion}
We introduce a new pre-trained representation for automatic sound effects library classification. Our task-specific representation outperforms OpenL3 on both UCS \& Non-UCS taxonomies across diverse datasets. Moreover, we show the effectiveness of cross-dataset training with Cross-Entropy loss over metric learning for this task. Results suggest that the quality and diversity of datasets are key to pre-training robust representations, supportive of preliminary experimental results where pre-training on increasing quantities of similar data yielded improvements with diminishing returns. Our contributions include that we are the first to conduct extensive deep learning experiments on UCS, experiment on relevant data for a current problem, and investigate an under-researched aspect of representation learning with cross-dataset training in the audio domain.

We believe cross-dataset training is a prerequisite for future work. Our current research does not leverage all datasets available to us nor the abundance of taxonomies they are comprised of given that many datasets did not have sufficient labels for training. To leverage these resources, methods to effectively work with large amounts of diverse data become increasingly essential. Selecting optimal cross-dataset training methods is fundamental before introducing new techniques, such as semi-supervised or self-supervised learning to work with datasets of varying label quality \cite{sohn2020fixmatch, gururani2021semi}. Accordingly, we would like to \begin{inparaenum}[(i)]
    \item look into the connection between multi-task learning and cross-dataset training to understand the specific advantages and drawbacks of these approaches and
    \item conduct experiments to assess the impact of increasing the amount of training data, observing the relationship of this with more complex network architectures and other pre-training methods \cite{music_sequence_representation}. \end{inparaenum} 
    Finally, we hope to deepen our understanding of UCS by investigating the non-uniformity of class labels between UCS-compliant datasets in this study, looking at the interpretability and disentanglement of our representations \cite{metric_vs_classification}, and exploring the concept of a generalized embedding from the perspective of taxonomy conversion \cite{text_to_text_transformer}.


\section{Acknowledgments}\label{sec:acknowledgments}
We would like to thank those who provided the data required to conduct this research as well as those who took the time to share their insights and software licenses for tools regarding sound search, query, and retrieval.\\

Alex Lane\footnote{Alex Lane, \textit{Resonant Rusty Door}. [Dataset]. Available: \href{https://www.alex-lane.com}{https://www.alex-lane.com}. [Accessed: September 15, 2021].}
\vspace{0.11cm}

All You Can Eat Audio\footnote{All You Can Eat Audio, \textit{It's A Plain Phone}. [Dataset]. Available:
\href{https://allyoucaneataudio.com/library-packs/it-is-a-plain-phone-p04}{https://allyoucaneataudio.com/library-packs/it-is-a-plain-phone-p04}. [Accessed: September 15, 2021].}
\footnote{All You Can Eat Audio, \textit{Cuff 'Em}. [Dataset]. Available: \href{https://allyoucaneataudio.com/library-packs/cuff-em}{https://allyoucaneataudio.com/library-packs/cuff-em}. [Accessed: September 15, 2021].}
\vspace{0.11cm}

Articulated Sounds\footnote{Articulated Sounds, \textit{Starter Pack}. [Dataset] Available: \href{https://articulatedsounds.com/audio-royalty-free-library/sfx/free-global-sample-pack}{https://articulatedsounds.com/audio-royalty-free-library/sfx/free-global-sample-pack}. [Accessed: September 15, 2021].}
\vspace{0.11cm}

Audio Shade\footnote{Audio Shade, \textit{Bravo Dramatic Audiences}. [Dataset]. Available: \href{https://audioshade.com/shop}{https://audioshade.com/shop}. [Accessed: September 15, 2021].} \footnote{Audio Shade, \textit{Brutality - Gore and Combat FX Toolkit}. [Dataset]. Available: \href{https://audioshade.com/shop}{https://audioshade.com/shop}. [Accessed: September 15, 2021].}
\vspace{0.11cm}

aXLSound\footnote{aXLSound, \textit{Planes - Takeoffs \& Landings}. [Dataset]. Available: \href{https://axlsound.com}{https://axlsound.com}. [Accessed: September 15, 2021].}
\vspace{0.11cm}

Big Sound Bank\footnote{Big Sound Bank, \textit{Big Sound Bank}. [Dataset]. Available: \href{https://bigsoundbank.com}{https://bigsoundbank.com}. [Accessed: September 15, 2021].}
\vspace{0.11cm}

BaseHead \footnote{\href{https://baseheadinc.com}{baseheadinc.com}, last accessed September 15, 2021.}
\vspace{0.11cm}

Bonson\footnote{Bonson, \textit{Wings}. [Dataset]. Available: \href{https://www.bonson.ca/product/wings}{https://www.bonson.ca/product/wings}. [Accessed: September 15, 2021].}
\footnote{Bonson, \textit{Moves}. [Dataset]. Available: \href{https://www.bonson.ca/product/moves}{https://www.bonson.ca/product/moves}. [Accessed: September 15, 2021].}
\vspace{0.11cm}

BOOM Library\footnote{BOOM Library, \textit{Cyber Weapons}. [Dataset]. Available: \href{https://www.boomlibrary.com/sound-effects/cyber-weapons}{https://www.boomlibrary.com/sound-effects/cyber-weapons}. [Accessed: September 15, 2021].}
\vspace{0.11cm}

Frick \& Traa\footnote{Frick \& Traa, \textit{City Bicycles}. [Dataset]. Available: \href{https://www.frickandtraa.com/webshop/sound-libraries}{https://www.frickandtraa.com/webshop/sound-libraries}. [Accessed: September 15, 2021].}
\pagebreak

Hzandbits\footnote{C. Hagelskjaer, "Interview with Christian Hagelskjaer," July 28, 2021.}
\vspace{0.11cm}

InspectorJ\footnote{InspectorJ, \textit{96 General Library}. [Dataset]. Available: \href{https://inspectorj.sellfy.store/p/96-general-library-bundle}{https://inspectorj.sellfy.store/p/96-general-library-bundle}. [Accessed: September 15, 2021].}
\vspace{0.11cm}

Kai Paquin\footnote{K. Paquin, \textit{Kailibrary}. [Dataset].}
\vspace{0.11cm}

KEDR Audio\footnote{KEDR Audio, \textit{Cinemaphone}. [Dataset]. Available: \href{https://www.asoundeffect.com/sound-library/cinemaphone-ringtones-notifications}{https://www.asoundeffect.com/sound-library/cinemaphone-ringtones-notifications}. [Accessed: September 15, 2021].}
\footnote{KEDR Audio, \textit{Kinetics: Construction Kit}. [Dataset]. Available: \href{https://www.asoundeffect.com/sound-library/kinetics-construction-kit}{https://www.asoundeffect.com/sound-library/kinetics-construction-kit}. [Accessed: September 15, 2021].} \footnote{KEDR Audio, \textit{Kinetics: Cinematic Tension}. [Dataset]. Available: \href{https://www.asoundeffect.com/sound-library/kinetics-cinematic-tension}{https://www.asoundeffect.com/sound-library/kinetics-cinematic-tension}. [Accessed: September 15, 2021].} \footnote{KEDR Audio, \textit{Radio Nomad}. [Dataset]. Available: \href{https://www.asoundeffect.com/sound-library/radio-nomad}{https://www.asoundeffect.com/sound-library/radio-nomad}. [Accessed: September 15, 2021].} \footnote{KEDR Audio, \textit{Vibrations}. [Dataset]. Available: \href{https://www.asoundeffect.com/sound-library/vibrations}{https://www.asoundeffect.com/sound-library/vibrations}. [Accessed: September 15, 2021].}
\vspace{0.11cm}

Krotos Audio\footnote{Krotos Audio, \textit{Ammo and Reloads}. [Dataset]. Available: \href{https://www.krotosaudio.com/products/ammo-reloads-sound-effects-library}{https://www.krotosaudio.com/products/ammo-reloads-sound-effects-library}. [Accessed: September 15, 2021].} \footnote{Krotos Audio, \textit{Mechanical}, 1. [Dataset]. Available: \href{https://www.krotosaudio.com/products/mechanical-sound-effects-library-vol-1}{https://www.krotosaudio.com/products/mechanical-sound-effects-library-vol-1}. [Accessed: September 15, 2021].}
\vspace{0.11cm}

Nikola Simikic\footnote{N. Simikic, \textit{Nikola Simikic Sound Library}. [Dataset].}
\vspace{0.11cm}

Penguin Grenade\footnote{Penguin Grenade, \textit{Arcs and Sparks}. [Dataset]. Available: \href{https://www.asoundeffect.com/sound-library/arcs-and-sparks}{https://www.asoundeffect.com/sound-library/arcs-and-sparks}. [Accessed: September 15, 2021].} \footnote{Penguin Grenade, \textit{Explosive Energy}. [Dataset]. Available: \href{https://www.asoundeffect.com/sound-library/explosive-energy}{https://www.asoundeffect.com/sound-library/explosive-energy}. [Accessed: September 15, 2021].} \footnote{Penguin Grenade, \textit{Holograms}. [Dataset]. Available: \href{https://www.asoundeffect.com/sound-library/holograms}{https://www.asoundeffect.com/sound-library/holograms}. [Accessed: September 15, 2021].}
\vspace{0.11cm}

Pro Sound Effects\footnote{Pro Sound Effects, \textit{CORE 2}, Pro. [Dataset]. Available: \href{https://www.prosoundeffects.com/core-2}{https://www.prosoundeffects.com/core-2}. [Accessed: September 15, 2021].}
\vspace{0.11cm}

Rick Allen Creative\footnote{R. Allen, "Interview with Rick Allen Creative," August 06, 2021.}
\vspace{0.11cm}

Sononym\footnote{B. Næsby, "Interview with Bjørn Næsby from Sononym," September 10, 2021.}
\vspace{0.11cm}

Sound Ideas\footnote{Sound Ideas, \textit{Sound Ideas Library}. [Dataset]. Available: \href{https://www.sound-ideas.com}{https://www.sound-ideas.com}. [Accessed: February 01, 2022].}
\vspace{0.11cm}

Soundly\footnote{Soundly, \textit{Soundly}. [Dataset]. Available: \href{https://www.soundly.com}{https://www.soundly.com}. [Accessed: February 01, 2022].}
\vspace{0.11cm}

Soundminer\footnote{\href{https://store.soundminer.com}{store.soundminer.com}, last accessed September 15, 2021.}
\vspace{0.11cm}

Storyblocks\footnote{\href{https://www.storyblocks.com}{storyblocks.com}, last accessed September 15, 2021.}
\vspace{0.11cm}

Tim Nielsen\footnote{T. Nielsen, \textit{Early Mother Communicator}. [Dataset].} \footnote{T. Nielsen, \textit{Ether}. [Dataset].} \footnote{T. Nielsen, \textit{Yellowstone}. [Dataset].} \footnote{T. Nielsen, \textit{Baltic}. [Dataset].}
\vspace{0.11cm}

Thomas Rex Beverly\footnote{T. R. Beverly, \textit{Maine Bundle}. [Dataset]. Available: \href{https://thomasrexbeverly.com/collections/sound-libraries/products/maine-bundle}{https://thomasrexbeverly.com/collections/sound-libraries/products/maine-bundle}. [Accessed: September 15, 2021].} \footnote{T. R. Beverly, \textit{Wild Animal Calls}. [Dataset]. Available: \href{https://thomasrexbeverly.com/products/wild-animal-calls}{https://thomasrexbeverly.com/products/wild-animal-calls}. [Accessed: September 15, 2021].} 
\vspace{0.11cm}

ZapSplat\footnote{A. McKinny, "Interview with Alan McKinny from ZapSplat," July 30, 2021.}

\end{document}